\documentclass{ws-procs9x6}

\def\lsim{\mathrel{\rlap{\lower3pt\hbox{$\sim$}}
    \raise2pt\hbox{$<$}}}

\begin{document}

\title{Vacuum \v{C}erenkov radiation in Maxwell-Chern-Simons electrodynamics}

\author{R.\ LEHNERT and R.\ POTTING}

\address{CENTRA, Departamento de F\'\i sica, FCT \\
Universidade do Algarve, \\ 
Campus de Gambelas, 8000 Faro, Portugal}

\maketitle

\abstracts{
We study the \v{C}erenkov effect in the context of the Maxwell-Chern-Simons 
(MCS) limit of the Standard Model Extension.
We present a method to determine the exact radiation rate for a point charge.
}

\section{The \v{C}erenkov effect in the MCS model}

In recent years the so-called Standard-Model Extension (SME)\cite{sme}
has provided a convenient framework for studying
minute Lorentz and CPT violations that may be
low-energy signatures for Planck-scale physics.\cite{cpt01}
In this work we will study a subsector of the SME describing
pure electrodynamics, where
Maxwell theory has been modified with Chern--Simons-like term in the
Lagrangian parametrized by dimensionful parameter $(k_{AF})^{\mu}$:
\begin{equation}
\mathcal{L}_{\rm MCS} = 
-{1\over4} F_{\mu\nu}F^{\mu\nu}
+(k_{AF})_{\mu}A_{\nu}\tilde{F}^{\mu\nu}
-A_{\mu}j^{\mu}.
\label{eq:lagrangian}
\end{equation}
The Chern--Simons term explicitly violates Lorentz invariance,
as well as PT and CPT invariance.
(For an explicit mechanism generating it see Ref.\ \refcite{vc}.)
We have explicitly included a coupling to an external current
$j^{\mu}$, which we take to satisfy $\partial_\mu j^\mu=0$.

As will become clear below, the inclusion of the $(k_{AF})_{\mu}$ term
results in a modification of the photon dispersion relation, with the
possibility of phase speeds smaller than the conventional speed of
light in vacuum $c$.
If realized in Nature, this opens up the possibility that ordinary
charged matter could move with a velocity exceeding the phase velocity
of radiation, and thus should emit \v{C}erenkov radiation \textit{in vacuum}.
This effect is well established experimentally and theoretically in
conventional macroscopic media.\cite{cer1}
Recently, some unexpected features have been encountered in
observations involving lead ions\cite{cern} and in exotic
condensed-matter systems.\cite{cms}
Some of these issues have been studied theoretically \cite{certheo}.

In this talk, we will present recent work by the present authors in which
vacuum \v{C}erenkov radiation was investigated in detail.\cite{us}
Our approach provides a new conceptual perspective on \v{C}erenkov
radiation, exploiting the fact that we have a fully relativistic
Lagrangian, that allows arbitrary observer Lorentz transformations.
In particular, going to the charge's rest frame turns out to simplify
the analysis.

\begin{figure}[ht]
\begin{center}
\includegraphics[width=0.6\hsize]{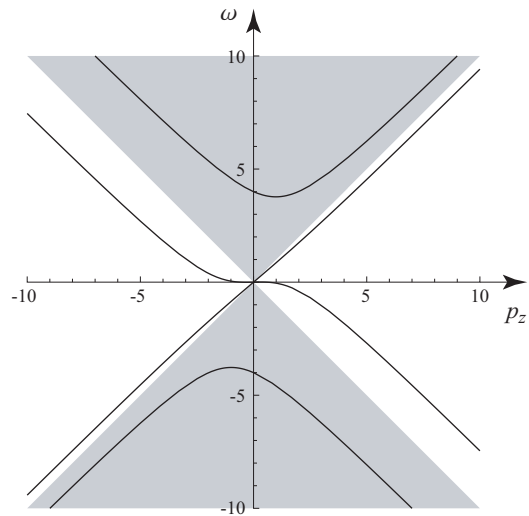}
\end{center}
\caption{Sample solution of the plane-wave dispersion relation.
The solid lines correspond to the exact roots.
The first-order solutions are shown as broken lines.
The shaded region represents
the interior of the $p^{\mu}$-space lightcone. \label{fig:dispersion}}
\end{figure}

The dispersion relation that follows from (\ref{eq:lagrangian})
is given by:
\begin{equation}
D(p^\mu)=p^4+4p^2k^2-4(p\cdot k)^2=0 .
\end{equation}
where $p^\mu=(\omega,\vec p)$ corresponds to the photon 4-momentum
and $k^\mu\equiv(k_{AF})^{\mu}$.
Generally, this dispersion relation includes time-like as well as
spacelike solutions for $p^\mu$. In figure \ref{fig:dispersion} the
case of space-like $k^\mu$ is depicted.
It can be shown that the spacelike and timelike  branches of the
dispersion relation correspond to deformed
elliptical polarizations.
At high momenta, they become left- and right circular polarizations.

In order to determine the rate of emission of \v{C}erenkov radiation,
it will be necessary to determine the solution of the equations
of motion in the presence of a charge, that is, with nonzero four-current.
The solution of the equation of motion that follows from lagrangian
(\ref{eq:lagrangian}) is:
\begin{equation}
A^{\mu}(x)=A^{\mu}_0(x)
+\int_{C_\omega}{d^4 p\over(2\pi)^4}\;\hat{G}^{\mu\nu}\hat{j\;}\!\!_{\nu}
\exp (-ip \cdot x),
\end{equation}
where $A^{\mu}_0(x)$ is any solution to the free equations of motion
(with $j^\mu=0$),
$\hat j_\nu$ is the Fourier transform of the current,
while the momentum space Green's function equals
\begin{equation}
\hat{G}^{\mu\nu} \equiv -{p^2 \eta^{\mu\nu}
+2i\varepsilon^{\mu\nu\rho\sigma}k_{\rho}p_{\sigma}
+4k^{\mu}k^{\nu}\over D(p^\mu)}+4\hat{G}^{\mu\nu}_0 ,
\end{equation}
where
\begin{equation}
\hat{G}^{\mu\nu}_0 \equiv
{(p\cdot k)
(p^{\mu}k^{\nu}+k^{\mu}p^{\nu})
-k^2 p^{\mu}p^{\nu}\over\big[D(p^\mu)\big]p^2},
\end{equation}
can be ignored as it
yields a total derivative upon contraction with a conserved current,
thus giving rise to a gauge artifact.
The integration contour $C_\omega$ has to be chosen judiciously to insure
retarded boundary conditions.

\section{Conditions for the emission of \v{C}erenkov radiation}
We will now determine the rate of emission of \v{C}erenkov
radiation by a pointlike charge.
As it turns out, the calculation is simplest in the rest frame of
the charge. As the current is time-independent in that frame, we have
for its Fourier transform
\begin{equation}
\hat j^\mu=2\pi\delta(\omega)\tilde j^\mu(\vec p)
\end{equation}
where $\tilde j^\mu(\vec p)$ is the Fourier transform in 3-space.
It follows 
\begin{equation}
A^\mu=\int{d^3\vec p\over(2\pi)^3}
{N^{\mu\nu}\tilde j_\nu(\vec p)\exp(i \vec p\cdot\vec r)\over D(0,\vec p)}
\label{eq:potential}
\end{equation}
with
\begin{equation}
N^{\mu\nu}(\vec{p}\hspace{1pt})\equiv\vec{p}^{\,2}\eta^{\mu\nu}
-2i\varepsilon^{\mu\nu\rho s}k_{\rho}p_{s}
-4k^{\mu}k^{\nu}.
\end{equation}
As the source is independent of time, the resulting electromagnetic
fields are expected to be stationary as well. 
Only spatial oscillations of the fields can occur.
This time independence suggests that the radiated energy shound be zero in
the rest frame of the charge.

Evaluating (\ref{eq:potential}), it is advantageous, as usual, to extend the
$|\vec p|$ integral to the complex plane, and use residue calculus.
It follows then directly that this integral yields a factor
\begin{equation}
\exp(i\vec p_0\cdot \vec r),
\end{equation}
where $\vec p_0$ satisfies the dispersion relation:
\begin{equation}
D(0,\vec p_0)=0.
\label{eq:dispersion}
\end{equation}
We conclude that a nonzero imaginary part of $p_0$ implies
exponential decay of the fields with increasing $r$, while a nonzero
real part corresponds an oscillatory behavior.
As transport of energy-momentum to infinity can only occur
in the presence of long-range fields,
it follows that we can expect
vacuum \v{C}erenkov radiation only if there are real four-momenta
$p^\mu=(0,\vec p)$
satisfying the plane wave dispersion relation in the charge's rest frame.

In a general frame, where charge's velocity is $\vec\beta'$,
the four-momentum $p^\mu=(0,\vec p)$ is transformed into
$(\vec\beta'\cdot\vec p',\vec p')$,
where
$\vec p'=\vec p+(\gamma-1)(\vec p\cdot\vec\beta')\vec\beta'/|\vec\beta'|^2$.
It follows that the phase velocity equals
\begin{equation}
c_{ph}'=|\vec\beta'\cdot\vec p'|/|\vec p'|\le|\vec\beta'|
\end{equation}
so that the velocity of the particle must exceed the phase velocity
of the waves. This corresponds exactly to the
conventional condition for emission of \v{C}erenkov radiation.

%
%

It is useful to consider the analogue of a boat in still water.
If the boat is in motion relative to the water,
a v-shaped wavefront appears.
For an observer on the boat, the wave pattern is stationary, while
for a general observer on the shore it oscillates
with decaying frequency (after the boat has passed).

\begin{figure}[ht]
\begin{center}
\includegraphics[width=0.8\hsize]{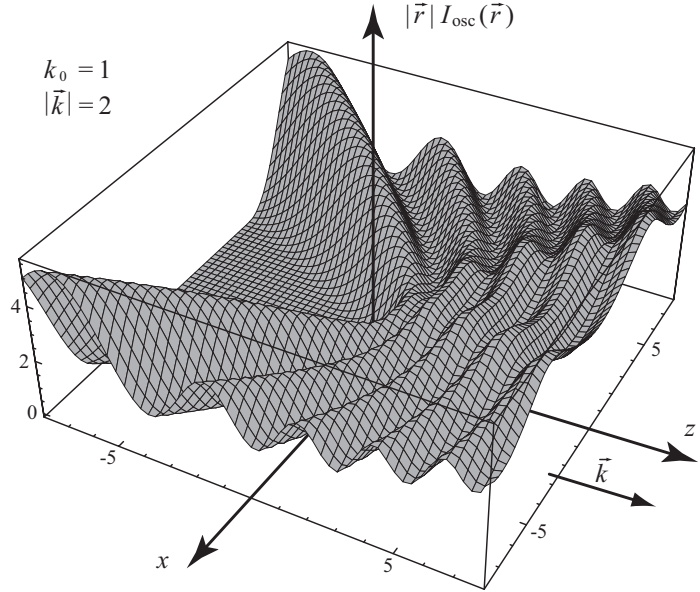}
\end{center}
\caption{General field pattern of a point charge resting at the origin.
The function 
$|\vec{r}\hspace{1pt}|I_{\rm osc}(\vec{r}\hspace{1pt})$ is shown
for $\vec{r}$ in the $xz$ plane 
with $\vec{k}$ along the $z$ direction. 
This function was evaluated
by an analytical  $|\vec{p}\hspace{1pt}|$-type integration followed 
by numerical angular integrations. 
Uninteresting nonoscillatory pieces 
$I_{\rm non}$ have been subtracted for clarity, 
so that only the oscillatory part $I_{\rm osc}\equiv I-I_{\rm non}$ 
contributes to this plot. 
The wave pattern is resemblant 
to that caused by a boat moving in water.
\label{fig:radiationplot}}
\end{figure}

Figure \ref{fig:radiationplot} depicts a quantity related to the potential as a function
of position, which clearly shows the nontrivial directional dependence
of the emitted waves.
Note that the MCS lagrangian implies a nontrivial
dispersion relation (\ref{eq:dispersion}).
Consequently, the direction of the \v{C}erenkov waves is frequency dependent,
resulting in the absence of a sharp shock-wave.

\section{Calculation of the emission rate}

The usual way to determining \v{C}erenkov rate involves 
integration of the $r^{-2}$ piece of Poynting vector over
the boundary surface of space at infinity.
However, this procedure is intractable in the present case,
because determination of the asymptotic fields turns out to be
difficult.

An alternative approach has been developed in Ref.\ \refcite{us}.
We start with the following expression for the energy-momentum tensor
\begin{equation}
\Theta^{\mu\nu}\equiv-F^{\mu\alpha}F^{\nu}{}_{ \alpha} 
+{1\over4}\eta^{\mu\nu}F^{\alpha\beta}F_{\alpha\beta} 
-k^{\nu}\tilde{F}^{\mu\alpha}A_{\alpha} 
\end{equation}
which obeys the conservation condition
\begin{equation}
\partial_\mu \Theta^{\mu\nu}=j_{\mu}F^{\mu\nu}. 
\end{equation}
Integrating this equation over 3-volume yields
\begin{equation}
\int_\sigma d\sigma^{l} \, \Theta_{l\nu}
= \int_V d^3\vec{r} \; j^{\mu}F_{\mu\nu}
- {\partial\over\partial t} \int\limits_{V}d^3\vec{r} \; \Theta_{0\nu}\; .
\label{eq:conservation_int}
\end{equation}
We now take static point charge source 
$J^\mu(\vec r)=(q\delta(\vec r),\vec0)$.
It follows from Eq.\ \ref{eq:conservation_int} that
\begin{equation}
\int_\sigma d\vec{\sigma}\cdot\vec{S} =0
\end{equation}
for the Poynting vector $\Theta_{l0}\equiv S_{l}=-S^{l}$,
so the net radiated energy is always zero in the charge's rest frame,
as anticipated.
There is, however, a nonzero rate of radiation of 3-momentum:
\begin{equation}
\dot{P}_s\equiv\int_{\sigma}d\sigma^{l}\Theta_{ls}
\end{equation}
which becomes
\begin{equation}
\dot {\vec{P}}=\int_{V}d^3\vec{r} J^{\mu}\vec{\nabla} A_{\mu}.
\end{equation}
Using the explicit (retarded) solution (\ref{eq:potential}) obtained for
$A_\mu$ one can calculate $\dot{\vec P}$ by regularizing the delta-function
defining the source, and performing the Fourier integral.
It follows that
\begin{equation}
\dot{\vec P}=-\hbox{sgn}(k_0){q^2\over4\pi}{k_0^4\over\vec k^2}\vec e_k.
\label{eq:rate}
\end{equation}
Note that, as a consequence, $\dot{\vec P}=0$ if $k_0=0$, that is,
there is no radiation in the rest frame unless $k_0$ is nonzero.

Transforming to general frame in which the charge has an arbitrary
velocity generally yields non-zero components for all components of 
$\dot P_\mu$
that depend on both $\vec \beta$ and $\vec k$.

\begin{figure}[ht]
\begin{center}
\includegraphics[width=0.5\hsize]{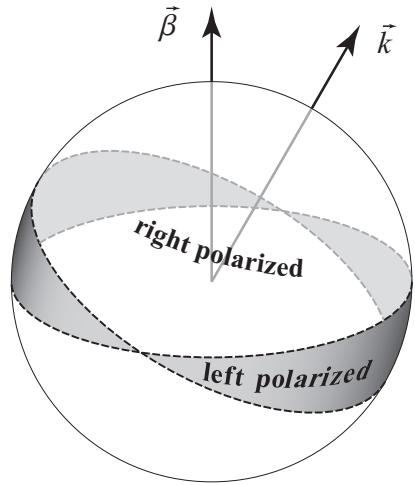}
\end{center}
\caption{Dependence of the polarization on direction.
For vectors $\vec{p}$
pointing in the clear (shaded) direction,
the associated waves are right (left) polarized.
The radiation exhibits linear polarization only
when $\vec{p}$ lies on one of the dashed lines.
Vacuum \v{C}erenkov radiation may not be emitted
into all directions.
The wave 4-vector $p^{\mu}=(\vec{\beta}\!\cdot\!\vec{p},\vec{p}\hspace{1pt})$
is further constrained by the dispersion relation.
\label{fig:polarization}}
\end{figure}

Figure \ref{fig:polarization} indicates the polarization of the
radiation as a function of the direction of the wave vector
$\vec p$ in relation to $\vec\beta$ and $\vec k$.

A natural question that presents itself is whether vacuum \v{C}erenkov
radiation might be observed.
As it turns out, in the laboratory frame the components of $k^{\mu}$
are observationally constrained by
$\mathcal{O} (k^{\mu})\lsim 10^{-42}\,$GeV.\cite{mcsclass}
The smallness of this bound implies that deviations of the photon phase
speed from $c$ are expected to be extremely small.
Taking this bound to be saturated,
it can be shown\cite{us} that a proton 
at the end of the observed cosmic-ray spectrum ($10^{20}\,$eV)
will only emit radiation of wavelengths larger than $1.2\times 10^{5}\,$m.
Conceivably, such radiation might be observable 
in high-energy astrophysical jets 
emitted in the direction of sight. 

\section{Back reaction on the charge}

Denoting the charge's 4-momentum by $Q^\mu$, momentum conservation yields
\begin{equation}
\dot Q^\mu = -\dot P^\mu(\vec\beta)
\end{equation}
It is possible to continue to consistently use the usual definition
$Q^\mu=m u^\mu$ so that one obtains the differential equtaion
\begin{equation}
-\dot P^\mu(\vec\beta)=m u^\mu(\vec\beta)
\label{eq:backreaction}
\end{equation}
where $\dot P^\mu(\vec\beta)$ has been determined in the previous
section (transforming formula (\ref{eq:rate}) to the appropriate frame).

For the important case of space-like $k^\mu$, this equation can be integrated
explicitly in the laboratory frame in which $k_0=0$, yielding the charge's
velocity as a function of time.\cite{us}
One can show that:
\begin{itemize}
\item The component $\beta_\bot$ normal to $\vec k$ is always constant in time;
\item The charge is always slowed down by \v Cerenkov radiation;
\item The characteristic time scale governing the time dependence is
given by $\tau =4\pi m/q^2\vec k^2\sqrt{1-\beta_\bot^2}$;
\item The trajectory is generally curved, with a characteristic scale
size $\tau\beta_\bot$.
\end{itemize}

One might speculate whether the slow-down effect of high-energy charges
might lead to an effective cut-off
in the cosmic-ray spectrum
for primary particles carrying an electric charge.
This idea has been raised in the literature
to place bounds on Lorentz breaking.
In the present model, however,
the energy-loss rate
is suppressed by two powers
of the (experimentally tightly bounded)
Lorentz-violating coefficient $k^{\mu}$.

\section{Phase space estimate}

While a full quantum field theory extension of the classical results
obtained above is beyond the scope of the current work, we will here present
a phase space estimate of the radiation rate and show it is consistent
with Eq.\ \ref{eq:rate}.

We start with the decay rate of a particle $a$ into two particles $b$ and $c$
in quantum fields theory:
\begin{equation}
d\Gamma={|\mathcal{M}_{a\rightarrow b,c}|^2\over2E_a}(2\pi)^4
\delta^{(4)}(p_a^{\mu}-p_b^{\mu}-p_c^{\mu})d\Pi_b d\Pi_c.
\label{eq:QFTdecayrate}
\end{equation}
Here $d\Pi_i$ ($i=a$, $b$, $c$) denote the phase-space elements.
We take $p_a^2=p_b^2=m^2$, corresponding to a mass $m$ particle
with a conventional Lorentz
invariant dispersion relation, while $c$ denotes photons with the
MCS dispersion relation.

It is possible to show the for light-like $k^\mu$, 
\begin{equation}
d\Pi_c={d^3\vec{p}_c\over(2\pi)^3 2|\vec{p}_c+{\rm sgn}(k^0)\vec{k}|}
\end{equation}
is observer-invariant for the space-like branches of the photon dispersion
relation.

For the amplitude we can take
\begin{equation}
\mathcal{M}_{a\rightarrow b,c}=qE_aM
\end{equation}
as the generic form of the amplitude, with $M$ a dimensionless function
of external momenta and the Lorentz-violating parameters. 

An order-of-magnitude estimate for expression (\ref{eq:QFTdecayrate})
can be worked out in the $m\to\infty$ limit,
yielding for the decay rate:
\begin{equation}
\dot{  \vec{P}}\simeq-{\rm sgn}(k^0)
{q^2\overline{|M|^2}\over8\pi}\vec{k}^{\,2}\vec{e}_k.
\end{equation}
Here $\overline{|M|^2}$ denotes a suitable angular average of $|M|^2$.
This result is in correspondence with classical result (\ref{eq:rate}).

\section{Conclusions}
We considered the possiblity of \v{C}erenkov radiation in the
Maxwell--Chern--Simons model, a particular limit of the SME.
We showed how the Lorentz-violating modification of the plane-wave dispersion
relation leads to the emission of radiation by moving charges.
Our novel approach exploited the fact that observer Lorentz invariance
always allows one to transform to the rest frame of the charge,
where the calculations are less complicated.
We investigated various properties of this radiation,
and obtained the exact (classical) rate of emission of radiation
by a point charge.
The possibility of detection of vacuum \v{C}erenkov radiation
in astrophysical context was considered,
with the conclusion that the tight observational
bounds on the $(k_{AF})^\mu$ parameter render any possible effect
highly suppressed.
We note that it would be interesting
to consider the dimensionless $k_{F}$ term in the SME:
some of its components 
are currently only bounded at the $10^{-9}$ level,\cite{cavexpt}
and a dynamical study paralleling the present one
could yield less suppressed rates.
We expect our methodology
to have applicability in more general cases including macroscopic media.

\end{document}